\documentclass[
    10pt,
    twocolumn,
    letterpaper,
    aps,
    prl,
    superscriptaddress,
    showpacs,
    amsmath,
    floatfix,
]{revtex4-1}

\usepackage{amssymb}
\usepackage{color}
\usepackage{graphicx}

\newcommand\UT{Department of Applied Physics, School of Engineering,\\ The University of Tokyo, 7-3-1 Hongo, Bunkyo-ku, Tokyo 113-8656, Japan}
\newcommand\UNSW{Centre for Quantum Computation and Communication Technology, School of Engineering and Information Technology, University of New South Wales Canberra, ACT 2610, Australia}

\begin{document}

\title{Real-Time Quadrature Measurement of a Single-Photon Wavepacket\\with Continuous Temporal-Mode-Matching}

\author{Hisashi Ogawa}
\affiliation{\UT}
\author{Hideaki Ohdan}
\affiliation{\UT}
\author{Kazunori Miyata}
\affiliation{\UT}
\author{Masahiro Taguchi}
\affiliation{\UT}
\author{\\Kenzo Makino}
\affiliation{\UT}
\author{Hidehiro Yonezawa}
\affiliation{\UNSW}
\author{Jun-ichi Yoshikawa}
\affiliation{\UT}
\author{Akira Furusawa}
\email{akiraf@ap.t.u-tokyo.ac.jp}
\affiliation{\UT}

\date{\today}

\begin{abstract}
Real-time controls based on quantum measurements are powerful tools for various quantum protocols.
However, their experimental realization have been limited by mode-mismatch between temporal mode of quadrature measurement and that heralded by photon detection.
Here, we demonstrate real-time quadrature measurement of a single-photon wavepacket induced by a photon detection, by utilizing continuous temporal-mode-matching between homodyne detection and an exponentially rising temporal mode.
Single photons in exponentially rising modes are also expected to be useful resources for interactions with other quantum systems.
\end{abstract}

\pacs{03.67.Lx, 42.50.Dv, 42.50.Ex, 42.65.-k}

\maketitle
Quantum measurement is a basic requirement for various quantum protocols.
In many of them, real-time acquisition of the measurement outcomes is beneficial or even required.
Real-time measurement enables real-time feedback or feedforward controls of quantum systems.
Measurement-based quantum computation (MBQC) \cite{GottesmanChuang, OneWayQC} is a typical example of such real-time usage of the measurement results, where the measurement results are fedforward to the next computational step.
In addition, quantum dynamics conditioned by measurements are often referred to as quantum trajectories or quantum filters \cite{Trajectory, BelavkinFilter, FilteringSinglePhoton}, and form essential parts of developing quantum control theories.
Based on them, basic cases of quantum controls, e.g., suppression of noises, have been tested with a variety of systems recently \cite{PhotonNumberFeedback, RabiOscillationFeedback, Photon-by-PhotonFeedback, MechanicalOscillatorFeedback}.

In order to put the quantum measurement-based technologies further, an important adaptation is to bridge different quantum systems or different quantum variables in the measurement system \cite{FilteringSinglePhoton, YanbeiPreparingaMechanicalOscillator, HybridQuantumDevices}.
As a general problem, one may want to connect different systems as a signal and probe.
Aside from the issues of connecting different systems, even when restricting us to the light field, there are two fundamental variables reflecting the wave-particle duality.
One is the continuous variable (CV) field quadrature amplitude $\hat x = (\hat a + \hat a^\dag)/\sqrt{2}$, and the other is the discrete variable (DV) photon number $\hat n = \hat a^\dag \hat a$, where $\hat a$ and $\hat a^\dag$ denote the field annihilation and creation operators, respectively.
CV-DV hybrid architectures have been identified to offer crucial advantages, ranging from deterministic and fault-tolerant MBQC \cite{GKP, MenicucciFaultTolarant, FurusawaBook, UlrikHybrid} to loophole-free Bell inequality test \cite{BellTestHybrid}.
Experimental technologies for the hybridization are also rapidly developing \cite{BelliniHybridEntanglement, JulienHybridEntanglement, TakedaQubitTele, SqueezingSinglePhoton}; however, combining them with real-time quantum-optical controls are facing difficulties.

An obstacle is a mode mismatch between CV-based homodyne detection and DV-based photon detection.
To be more precise, homodyne detection is sensitive to vacuum fluctuation, while photon detection is not.
Therefore, filtration of the field quadrature, i.e. integration with an appropriate weight function $f(t)$ as $\hat X=\int f(t)\hat x(t)dt$ is essential in order to remove irrelevant vacuum noises and obtain meaningful quadrature information matching with the field mode heralded by photon detections.
At this point, quadrature measurement often fails to be in real time.
That is critical for quantum feedback or feedforward controls since any latency can result in decoherence of the systems.
For example, previous experiments of CV-DV hybrid MBQC have been limited to Gaussian operations \cite{TakedaQubitTele, SqueezingSinglePhoton}, where the feedforwards are linear, and thus filtration can be done after the feedforwards.
In order to complete the universal operations of the hybrid MBQC \cite{GKP, MenicucciFaultTolarant, FurusawaBook, UlrikHybrid}, one non-Gaussian operation is required in addition to the Gaussian operations \cite{QCoverCV, UniversalCVQC, PetrCubicNonlinearity}.
In non-Gaussian operations, the feedforwards become nonlinear \cite{DSG}, and filtration must be finished before the nonlinear feedforwards since they are not commutable.
Thus, there has been growing demand for the real-time quadrature measurement of light fields in DV-based states.

In this Letter, we demonstrate real-time quadrature measurement of a single-photon wavepacket by creating the state in an exponentially rising field mode.
Quadrature values of the single-photon wavepackets are obtained without additional latency in the filtration, thanks to the continuous temporal-mode-matching between the homodyne detection and the field mode heralded by a photon detection, as explained later.
Our experiment is based on a specially designed singly-resonant optical parametric oscillator (OPO), which we call an asymmetric OPO.
Purity of the created single photon is so high that strongly non-Gaussian behavior of the continuous quadrature signals is obtained in real-time on an oscilloscope without any postprocessing of the data or compensation for losses.

Single photons in exponentially rising modes are not only good for bridging different variables of a light field itself, but also for bridging different quantum systems.
For example, they can perfectly excite two-level atoms since they are the time reversals of photons emitted by exponentially decaying process of the atoms \cite{PerfectExcitation}.
More generally, the exponentially rising modes are known to be the most efficient modes for interactions with such systems as optomechanical oscillators, superconducting circuits, etc.\@ \cite{PerfectExcitation, YanbeiPreparingaMechanicalOscillator,  ZeroDynamics, CatchingTime-ReversedMicrowave}.
Therefore, our single photon in an exponentially rising mode can also be utilized as single-photon probe fields in quantum trajectories and filters \cite{FilteringSinglePhoton}, or as carriers in quantum networks \cite{QuantumStateTransfer, KimbleQuantumInternet, DLCZ}.

There are two possible ways to achieve the weighted integration of a field quadrature.
One is to tailor the envelope of local oscillator (LO) for homodyne detection to $f(t)$ and then integrate the homodyne signal (optical filtration).
The other is to continuously measure the field quadrature with continuous-wave (CW) LO and electrically perform weighted integration (electrical filtration).
The optical filtrations are natural choices for pulsed-laser-based experiments \cite{LvovskySinglePhotonHomodyne, BelliniAdaptiveDetection}.
On the other hand, the electrical filtrations are used in many successful experiments with CW-pumped photon sources \cite{JonasCat, JonasBrightSinglePhoton, LvovskyNarrowBandPhoton, LeeCatTele}, including some CV-DV hybrid protocols \cite{JulienHybridEntanglement, TakedaQubitTele, SqueezingSinglePhoton}.
The electrical filtrations are suitable for such random photon sources since they continuously measure the field quadrature.
The drawback of electrical filtrations is that they have used digital computers for the complicated integration, which is far from real time.
However, when the target field mode is an exponentially rising mode around time $t_0$
\begin{equation}
f_\text{rise}(t ; t_0) = \sqrt{\gamma}e^{\gamma (t-t_0)/2}\Theta(t_0-t),\label{eq:exprise}
\end{equation}
where $\Theta(t)$ is the Heaviside step function, the filtration corresponds to a simple first-order low-pass filter (LPF).
By adding an LPF to a CW-LO-based homodyne detection, we can obtain the filtrated quadrature information
\begin{equation}
\hat X(t_\text{m}) = \int \sqrt{\gamma_\text m}e^{\gamma_\text m (t-t_\text m)/2}\Theta(t_\text m-t)\hat x(t)dt, \label{eq:measure}
\end{equation}
which is \textit{continuously matched} with the exponentially rising mode.
At $t_\text m = t_0$, the outcome corresponds to the quadrature value of the target field mode $f_\text{rise}(t;t_0)$.
Such a simple LPF can consist of only passive electrical components, or a finite-bandwidth homodyne detector itself can work as the LPF \cite{KimblePhotoDetectionProcesses}.
Therefore, there is no additional latency in the filtration, and we can overcome the drawback of electrical filtrations.
The continuous temporal-mode-matching can be extended to more general wavepackets as long as they can be matched with simple filters.

\begin{figure}
  \begin{center}
   \includegraphics[scale = 0.6]{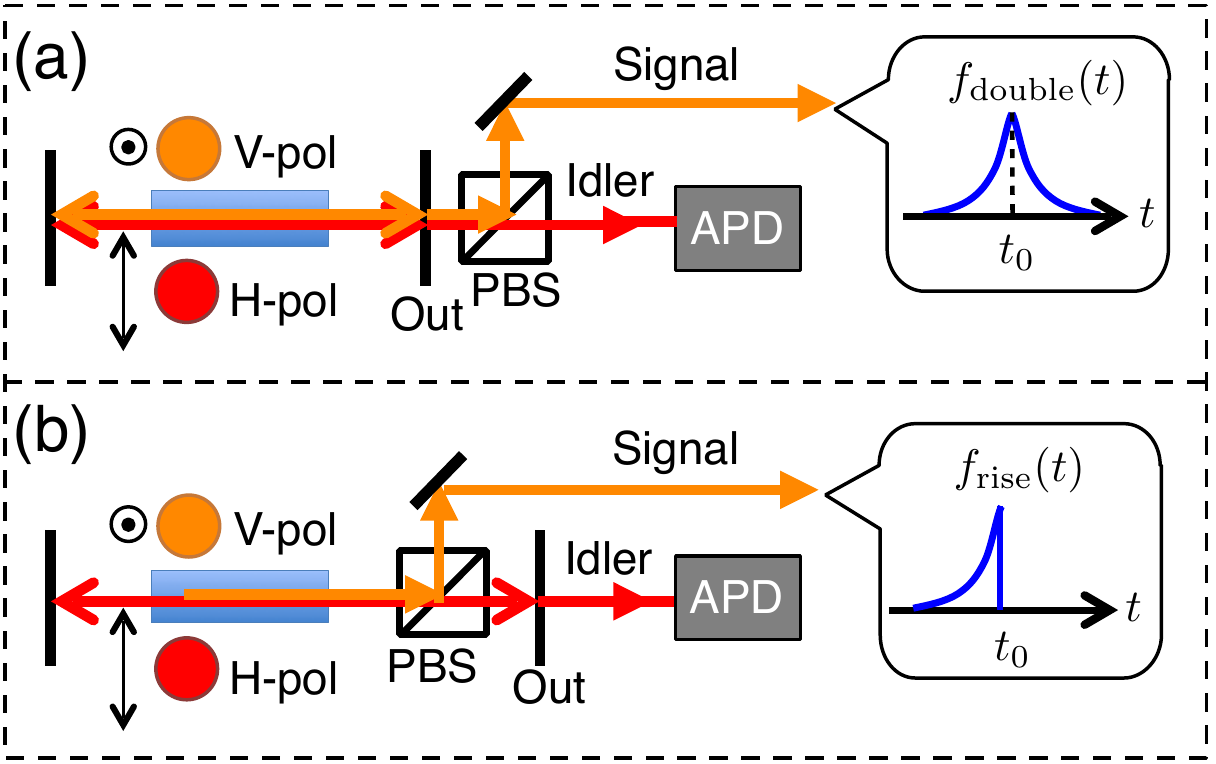}
  \end{center}
\caption{\label{Fig1}(color online) Single photon generation methods with (a) conventional doubly resonant OPO and (b) our asymmetric OPO; Out, outcoupling mirror; APD, avalanche photodiode for idler-photon detection.}
\end{figure}
We create a single photon with a heralding scheme, where photon pairs are probabilistically generated and detection of one photon (idler photon) heralds the other photon (signal photon).
A conventional doubly-resonant OPO [Fig.~\ref{Fig1}(a)] creates the signal photon in a double exponential mode $f_\text{double}(t;t_0) = \sqrt{\gamma/2} \exp(-\gamma |t-t_0|/2)$, where $t_0$ is the heralding timing and $\gamma$ corresponds to the bandwidth of the OPO \cite{NielsenSinglePhoton}.
This is because both signal and idler photons are resonant with the OPO, and leak from the OPO with the same decay rate $\gamma$.
On the other hand, our asymmetric OPO [Fig.~\ref{Fig1}(b)] introduces asymmetry in the decay rate of the photon pairs by a polarization beam splitter (PBS) inside the cavity.
Here, vertically-polarized (V-pol) signal photon is ejected by the PBS, and thus, quickly decays after the detection of horizontally-polarized (H-pol) idler photon.
As the result, the wavepacket of the signal photon is approximated by $f_\text{rise}(t;t_0)$.
Creating single photons in exponentially rising modes have been tested mainly with atomic ensembles  \cite{LvovskyComplete, CoherentControl, GulatiGenerationExprise, EfficientlyLoading, Reversing}.
However, these previous experiments have reported nearly 50\% loss or have been verified just by the property of antibunching which is insensitive to the loss.
Compared with them, the asymmetric OPO has several useful features.
Transverse modes of the photons are cleaned by the OPO so that we can easily get high interference visibility with the LO or other beams.
Thus, we can obtain the information of the state with high purity.
Moreover, our method can be extended to other wavelengths by using appropriate nonlinear crystals, and $\gamma$ can be changed by the bandwidth of the OPO.
Adjustability of these parameters suggests that our method can create single photons matching with a variety of other systems \cite{PerfectExcitation, YanbeiPreparingaMechanicalOscillator, ZeroDynamics}.
We also note that we can create superposition of Fock states by adding displacement to the idler photons \cite{Lvovsky2photon, YukawaSuperposition}.

\begin{figure}
  \begin{center}
   \includegraphics[scale = 0.85]{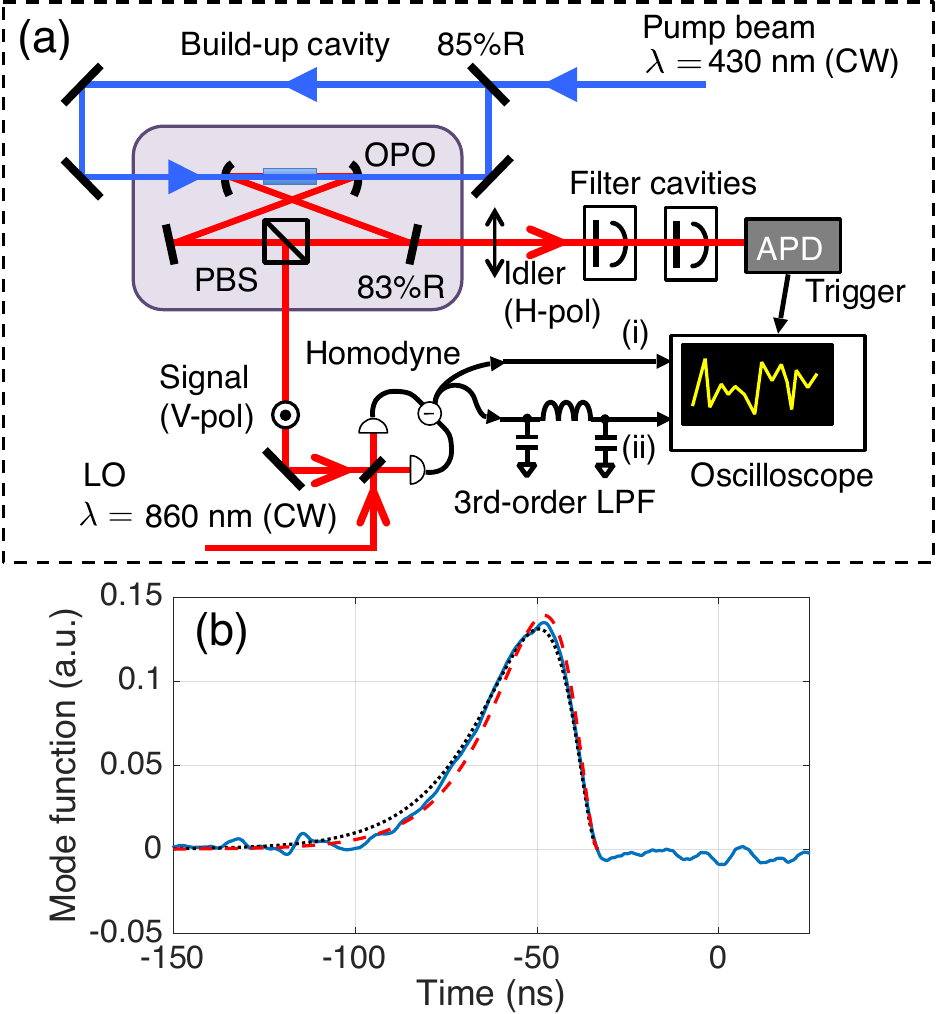}
  \end{center}
\caption{\label{Fig2}(color online) Experimental setup and the result of preliminary experiment. (a) Experimental setup; $\lambda$, wavelength; R, reflectivity. (b) Solid trace: estimated temporal mode of the single-photon wavepacket. Dashed curve: weighting function of the 3rd-order LPF in the signal line (ii) in Fig.~\ref{Fig2}(a). Dotted curve: temporal mode calculated by substituting the bandwidths of the cavities into Eq.~\eqref{eq:3rdLPF}.}
\end{figure}

The experimental setup is shown in Fig.~\ref{Fig2}(a).
We use a continuous-wave titanium sapphire laser operating at wavelength of 860~nm.
The OPO, containing a periodically-poled KTiOPO$_4$ crystal with type-II phase-matching, is pumped with a 430~nm CW beam which is enhances by a build-up cavity.
The build-up cavity also improves the stability of the transverse mode of created signal photon by stabilizing the optical path of the pump beam.
Since the OPO is pumped with a CW beam, the single-photon wavepacket is created at random timing.
Bandwidth of the OPO is designed to be about 11~MHz of half width at half maximum (HWHM).
The signal of the homodyne detector is recorded with an oscilloscope triggered by the H-pol idler-photon detection at an avalanche photodiode (APD).
Two filter cavities (FCs) which remove photons in irrelevant frequencies are employed in the idler line.
Their bandwidths are designed to be about 19~MHz and 36~MHz of HWHM, respectively.
Since the FCs work as additional Lorentzian filters, the temporal mode of the heralded signal photon $f_\text{rise}^\text{exp}(t;t_0)$ becomes linear combination of three exponentially rising modes as
\begin{equation}
f_\text{rise}^\text{exp}(t;t_0) \propto \sum_{n=1}^3c_ne^{\gamma_n (t-t_0)/2}\Theta(t_0-t),\label{eq:3rdLPF}
\end{equation}
where $\gamma_\text 1, \gamma_2, \gamma_3$ correspond to the bandwidths of the cavities, $c_1 = 1/(\gamma_2-\gamma_1)(\gamma_3-\gamma_1)$ and $c_2, c_3$ are its cyclic permutations.
In this case, the filtration corresponds to a 3rd-order LPF.

For the verification, we compare the conventional digital postprocessing method and the real-time method by simultaneously recording the unfiltered and filtered homodyne signal [signal line (i) and (ii) in Fig.~\ref{Fig2}(a), respectively].
They are recorded during 500~ns around each heralding signal with a sampling rate of 2.5~GHz.
The unfiltered signal is analyzed after collecting a series of data, using the same manner as \cite{LvovskyNarrowBandPhoton, JulienTemporalMode}.
On the other hand, the 3rd-order LPF for the real-time measurement consists of only passive electrical components (two chipped capacitors and one chipped inductor as in Fig.~\ref{Fig2}(a)), and its circuit length is about 5~mm.
There is no latency in the filtration except for the propagation time in the electric cables.
The event rate of photon detection is about 1,800 per second, and we collect the data of 18,491 events.
Bandwidth of the homodyne detector is 100~MHz, which is much broader than those of the cavities so that it does not distort the detected wavepacket.
Since single photons are phase insensitive, we scanned the LO phase at about 100~Hz while recording the data.

The solid trace in Fig.~\ref{Fig2}(b) is the temporal mode of the single-photon wavepacket estimated from the principle component analysis \cite{LvovskyNarrowBandPhoton, JulienTemporalMode} of the unfiltered homodyne signal.
It clearly shows an exponentially rising and suddenly falling feature.
Fig.~\ref{Fig2}(b) also shows the theoretical curve calculated by substituting the bandwidths of the cavities into Eq.~\eqref{eq:3rdLPF} and the weighting function of the 3rd-order LPF in the signal line (ii) in Fig.~\ref{Fig2}(a).
The mode-match between the single-photon wavepacket and these curves are over 99\%.
Although the performance of the real-time quadrature measurement does not depend on whether the LPF is 1st-order or 3rd-order, it should be noted that we can make the wavepacket closer to single exponentially rising mode as Eq.~\eqref{eq:exprise} by using more broad-band FCs.

\begin{figure}
  \begin{center}
  \includegraphics[scale = 0.19]{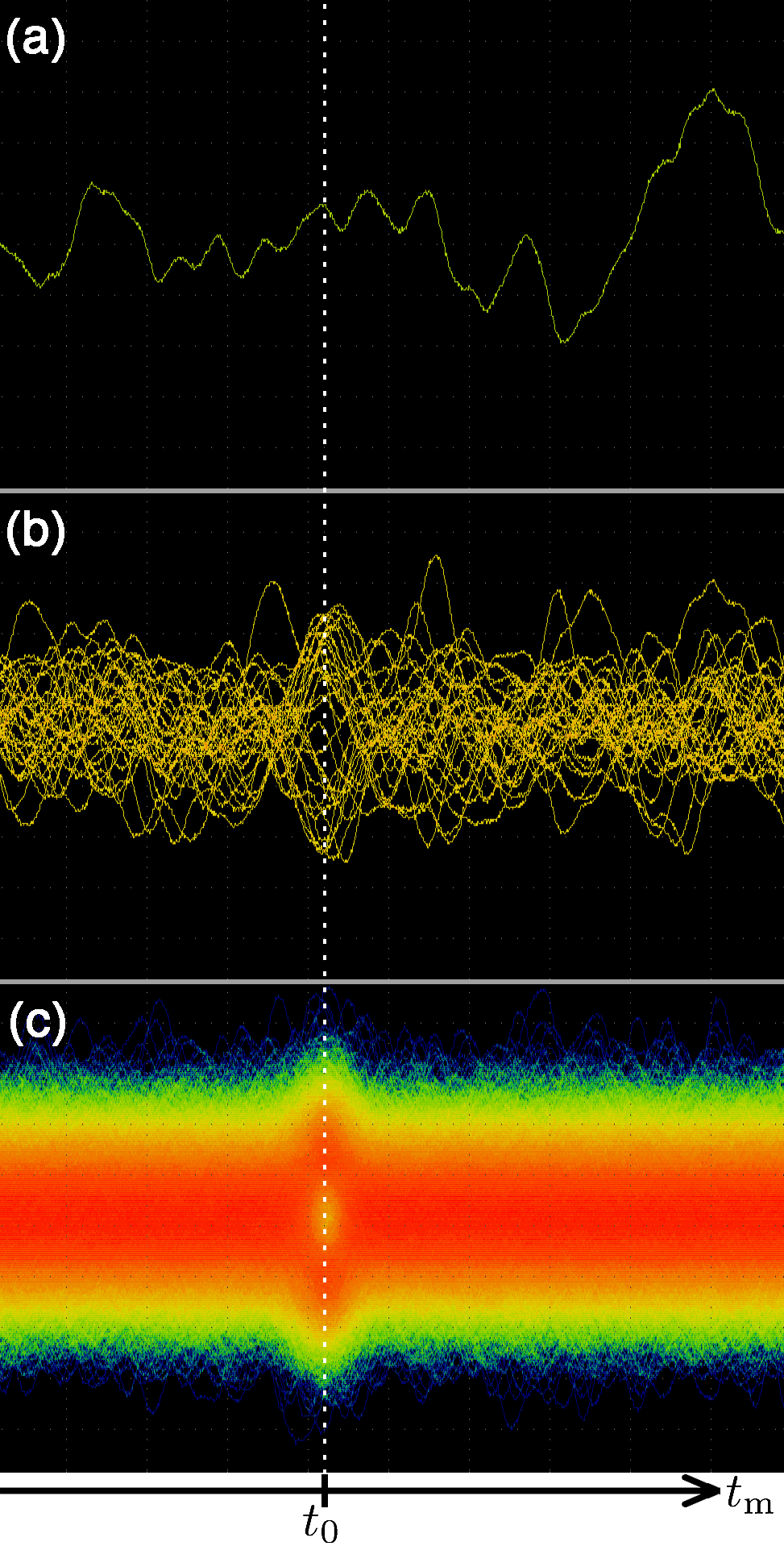}
  \end{center}
\caption{\label{Fig3}(color online) Screen captures of the oscilloscope displaying the homodyne signal after the LPF \cite{supplemental}. They are recorded 500~ns around each heralding event. (a) Single trace of the signal. (b) Overlaid 32 traces. (c) Overlaid 18,491 traces. $t_\text m$ is measurement time, and vertical dotted line indicates the heralding signal at time $t_\text m = t_0$. Histogram at each time is visualized by colors changing from blue to yellow to red as the sample density increaces.}
\end{figure}
Figure~\ref{Fig3} shows the screen captures of the oscilloscope displaying the  traces of the filtered homodyne signals without any digital postprocess or compensation for losses.
Here, the quadrature information correlated with the idler field $\hat a_\text i(t_m)$ is continuously monitored as the real-time signal at $t_\text m$.
The vertical dotted line in Fig.~\ref{Fig3} indicates the timing of the heralding signals, which corresponds to $t_0$ in Eq.~\eqref{eq:3rdLPF}.
Each signal on the white dotted line corresponds to the quadrature of the single-photon wavepacket.
As the number of overlaid traces increases, a non-Gaussian dip gradually appears on the white dotted line \cite{supplemental}, which is characteristic of quadrature distributions of single photons.
Continuous change of the distribution with respect to $t_\text m$ in Fig.~\ref{Fig3}(c) reflects the overlap between the single-photon wavepacket $f_\text{rise}^\text{exp}(t;t_0)$ and the mode filtrated in real-time $f_\text{rise}^\text{exp}(t;t_\text m)$.
Since they are in the same shape, the overlap depends on $|t_0 - t_\text m|$.
That is why we can see the symmetric distribution about $t_\text 0$.
Observing the particle nature of light in real-time by the wave-based measurements is in contrast to interference-diffraction experiments where wave nature is observed by particle-based measurements, e.g., \cite{TwoPhotonGhost}.

\begin{figure}
  \begin{center}
   \includegraphics[scale = 0.66]{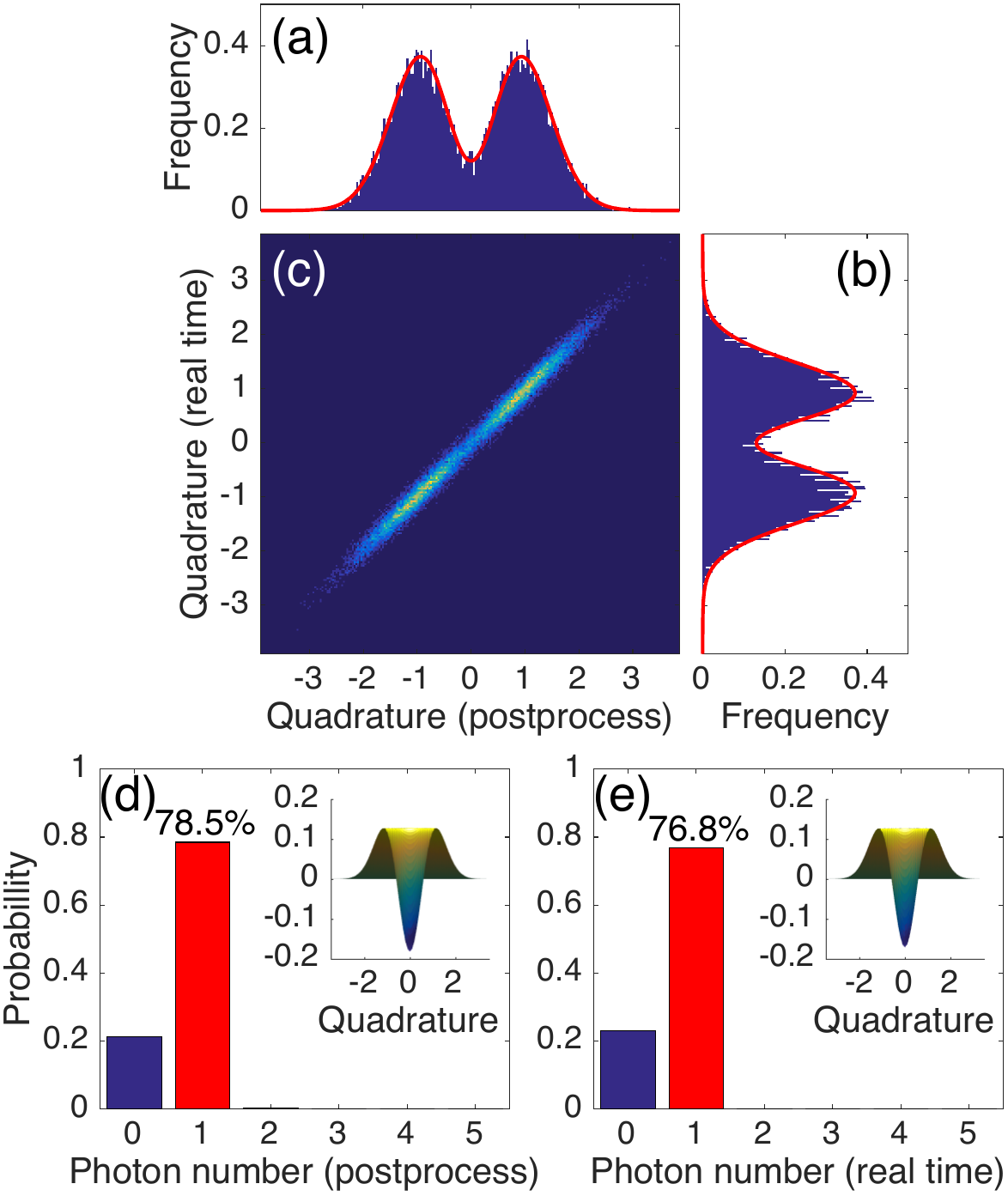}
  \end{center}
\caption{\label{Fig4}(color online) Quadrature distributions and estimated diagonal elements of density matrices. (a) and (b) Quadrature distributions obtained by the postprocessing method and the real-time method, respectively. Solid curves are their fitting curves. (c) Joint distribution of the quadratures. (d) and (e) Diagonal elements of density matrices estimated from (a) and (b), respectively. Sectional side view of their Wigner functions cutting through the phase-space origin are displayed as insets.}
\end{figure}

A comparison of quadrature distributions obtained by the real-time method and the postprocessing method are shown in Fig.~\ref{Fig4}.
Figure~\ref{Fig4}(a) is the quadrature distribution obtained by postprocessing of the recorded unfiltered signal.
On the other hand, Fig.~\ref{Fig4}(b) is the quadrature distribution of the real-time measurement at $t_\text m = t_0$, which corresponds to the distribution of the signals on the dotted line in Fig.~\ref{Fig3}(c).
Figure~\ref{Fig4}(c) is the joint distribution of the quadratures, and it clearly shows strong correlation between the quadratures obtained by the two methods.
Correlation coefficient calculated from the distribution is over 0.99, which means almost the same quadrature values are obtained for each event.
We also estimate the diagonal elements of their density matrices from the quadrature distributions using the maximum-likelihood method \cite{Maxlik}.
The error is estimated by the bootstrap method.
The single-photon components of the results of postprocessing method and the real-time method are $78.5 \pm 0.7~\%$ and $76.8 \pm 0.6~\%$, respectively [Figs.~\ref{Fig4} (d) and (e)].
Wigner functions reconstructed from the density matrices are also shown in the insets.
They have negative dips at the origin with values of $-0.181 \pm 0.004$ and $-0.170 \pm 0.004$, respectively.
These results clearly show that the highly pure single photon is measured with high temporal-mode-matching.

In summary, we have demonstrated real-time quadrature measurement of a single-photon wavepacket, which is created in an exponentially rising mode, with the continuous temporal-mode-matching.
We succeeded in obtaining the quadrature values of the randomly arriving single-photon wavepacket with no additional latency.
From the quadrature distributions, we confirmed the high purity of the created single photons, and the high temporal-mode-matching between the homodyne detection and the single-photon wavepacket.
Our method is a powerful tool for the CV-DV hybrid quantum protocols which involves real-time controls.
We also expect the highly pure single photon in an exponentially rising mode will be a useful resource for connecting different systems, and will stimulate a variety of studies, from quantum trajectories and filters to quantum networks.

We thank John E. Gough, Matthew R. James, Hendra I. Nurdin, Naoki Yamamoto,  Petr Marek, and Radim Filip for valuable and fruitful discussions.
This work was partly supported by PDIS, GIA, APSA commissioned by the MEXT of Japan, ASCR-JSPS, the Australian Research Council, Grant No.\@ CE1101027, CREST of JST, and the SCOPE program of the MIC of Japan.
H.\@ Og.\@ and K.\@ Mi.\@ acknowledge financial support from ALPS.

\bibliography{realtime}

\begin{thebibliography}{49}%
\makeatletter
\providecommand \@ifxundefined [1]{%
 \@ifx{#1\undefined}
}%
\providecommand \@ifnum [1]{%
 \ifnum #1\expandafter \@firstoftwo
 \else \expandafter \@secondoftwo
 \fi
}%
\providecommand \@ifx [1]{%
 \ifx #1\expandafter \@firstoftwo
 \else \expandafter \@secondoftwo
 \fi
}%
\providecommand \natexlab [1]{#1}%
\providecommand \enquote  [1]{``#1''}%
\providecommand \bibnamefont  [1]{#1}%
\providecommand \bibfnamefont [1]{#1}%
\providecommand \citenamefont [1]{#1}%
\providecommand \href@noop [0]{\@secondoftwo}%
\providecommand \href [0]{\begingroup \@sanitize@url \@href}%
\providecommand \@href[1]{\@@startlink{#1}\@@href}%
\providecommand \@@href[1]{\endgroup#1\@@endlink}%
\providecommand \@sanitize@url [0]{\catcode `\\12\catcode `\$12\catcode
  `\&12\catcode `\#12\catcode `\^12\catcode `\_12\catcode `\%12\relax}%
\providecommand \@@startlink[1]{}%
\providecommand \@@endlink[0]{}%
\providecommand \url  [0]{\begingroup\@sanitize@url \@url }%
\providecommand \@url [1]{\endgroup\@href {#1}{\urlprefix }}%
\providecommand \urlprefix  [0]{URL }%
\providecommand \Eprint [0]{\href }%
\providecommand \doibase [0]{http://dx.doi.org/}%
\providecommand \selectlanguage [0]{\@gobble}%
\providecommand \bibinfo  [0]{\@secondoftwo}%
\providecommand \bibfield  [0]{\@secondoftwo}%
\providecommand \translation [1]{[#1]}%
\providecommand \BibitemOpen [0]{}%
\providecommand \bibitemStop [0]{}%
\providecommand \bibitemNoStop [0]{.\EOS\space}%
\providecommand \EOS [0]{\spacefactor3000\relax}%
\providecommand \BibitemShut  [1]{\csname bibitem#1\endcsname}%
\let\auto@bib@innerbib\@empty
\bibitem [{\citenamefont {Gottesman}\ and\ \citenamefont
  {Chuang}(1999)}]{GottesmanChuang}%
  \BibitemOpen
  \bibfield  {author} {\bibinfo {author} {\bibfnamefont {D.}~\bibnamefont
  {Gottesman}}\ and\ \bibinfo {author} {\bibfnamefont {I.~L.}\ \bibnamefont
  {Chuang}},\ }\href {http://dx.doi.org/10.1038/46503} {\bibfield  {journal}
  {\bibinfo  {journal} {Nature}\ }\textbf {\bibinfo {volume} {402}},\ \bibinfo
  {pages} {390} (\bibinfo {year} {1999})}\BibitemShut {NoStop}%
\bibitem [{\citenamefont {Raussendorf}\ and\ \citenamefont
  {Briegel}(2001)}]{OneWayQC}%
  \BibitemOpen
  \bibfield  {author} {\bibinfo {author} {\bibfnamefont {R.}~\bibnamefont
  {Raussendorf}}\ and\ \bibinfo {author} {\bibfnamefont {H.~J.}\ \bibnamefont
  {Briegel}},\ }\href {\doibase 10.1103/PhysRevLett.86.5188} {\bibfield
  {journal} {\bibinfo  {journal} {Phys. Rev. Lett.}\ }\textbf {\bibinfo
  {volume} {86}},\ \bibinfo {pages} {5188} (\bibinfo {year}
  {2001})}\BibitemShut {NoStop}%
\bibitem [{\citenamefont {Carmichael}(1993)}]{Trajectory}%
  \BibitemOpen
  \bibfield  {author} {\bibinfo {author} {\bibfnamefont {H.}~\bibnamefont
  {Carmichael}},\ }\href@noop {} {\emph {\bibinfo {title} {An Open Systems
  Approach to Quantum Optics}}}\ (\bibinfo  {publisher} {Springer},\ \bibinfo
  {year} {1993})\BibitemShut {NoStop}%
\bibitem [{\citenamefont {Barchielli}\ and\ \citenamefont
  {Belavkin}(1991)}]{BelavkinFilter}%
  \BibitemOpen
  \bibfield  {author} {\bibinfo {author} {\bibfnamefont {A.}~\bibnamefont
  {Barchielli}}\ and\ \bibinfo {author} {\bibfnamefont {V.~P.}\ \bibnamefont
  {Belavkin}},\ }\href {http://stacks.iop.org/0305-4470/24/i=7/a=022}
  {\bibfield  {journal} {\bibinfo  {journal} {Journal of Physics A:
  Mathematical and General}\ }\textbf {\bibinfo {volume} {24}},\ \bibinfo
  {pages} {1495} (\bibinfo {year} {1991})}\BibitemShut {NoStop}%
\bibitem [{\citenamefont {Gough}\ \emph {et~al.}(2012)\citenamefont {Gough},
  \citenamefont {James}, \citenamefont {Nurdin},\ and\ \citenamefont
  {Combes}}]{FilteringSinglePhoton}%
  \BibitemOpen
  \bibfield  {author} {\bibinfo {author} {\bibfnamefont {J.~E.}\ \bibnamefont
  {Gough}}, \bibinfo {author} {\bibfnamefont {M.~R.}\ \bibnamefont {James}},
  \bibinfo {author} {\bibfnamefont {H.~I.}\ \bibnamefont {Nurdin}}, \ and\
  \bibinfo {author} {\bibfnamefont {J.}~\bibnamefont {Combes}},\ }\href
  {\doibase 10.1103/PhysRevA.86.043819} {\bibfield  {journal} {\bibinfo
  {journal} {Phys. Rev. A}\ }\textbf {\bibinfo {volume} {86}},\ \bibinfo
  {pages} {043819} (\bibinfo {year} {2012})}\BibitemShut {NoStop}%
\bibitem [{\citenamefont {Sayrin}\ \emph {et~al.}(2011)\citenamefont {Sayrin},
  \citenamefont {Dotsenko}, \citenamefont {Zhou}, \citenamefont {Peaudecerf},
  \citenamefont {Rybarczyk}, \citenamefont {Gleyzes}, \citenamefont {Rouchon},
  \citenamefont {Mirrahimi}, \citenamefont {Amini}, \citenamefont {Brune},
  \citenamefont {Raimond},\ and\ \citenamefont
  {Haroche}}]{PhotonNumberFeedback}%
  \BibitemOpen
  \bibfield  {author} {\bibinfo {author} {\bibfnamefont {C.}~\bibnamefont
  {Sayrin}}, \bibinfo {author} {\bibfnamefont {I.}~\bibnamefont {Dotsenko}},
  \bibinfo {author} {\bibfnamefont {X.}~\bibnamefont {Zhou}}, \bibinfo {author}
  {\bibfnamefont {B.}~\bibnamefont {Peaudecerf}}, \bibinfo {author}
  {\bibfnamefont {T.}~\bibnamefont {Rybarczyk}}, \bibinfo {author}
  {\bibfnamefont {S.}~\bibnamefont {Gleyzes}}, \bibinfo {author} {\bibfnamefont
  {P.}~\bibnamefont {Rouchon}}, \bibinfo {author} {\bibfnamefont
  {M.}~\bibnamefont {Mirrahimi}}, \bibinfo {author} {\bibfnamefont
  {H.}~\bibnamefont {Amini}}, \bibinfo {author} {\bibfnamefont
  {M.}~\bibnamefont {Brune}}, \bibinfo {author} {\bibfnamefont {J.-M.}\
  \bibnamefont {Raimond}}, \ and\ \bibinfo {author} {\bibfnamefont
  {S.}~\bibnamefont {Haroche}},\ }\href {http://dx.doi.org/10.1038/nature10376}
  {\bibfield  {journal} {\bibinfo  {journal} {Nature}\ }\textbf {\bibinfo
  {volume} {477}},\ \bibinfo {pages} {73} (\bibinfo {year} {2011})}\BibitemShut
  {NoStop}%
\bibitem [{\citenamefont {Vijay}\ \emph {et~al.}(2012)\citenamefont {Vijay},
  \citenamefont {Macklin}, \citenamefont {Slichter}, \citenamefont {Weber},
  \citenamefont {Murch}, \citenamefont {Naik}, \citenamefont {Korotkov},\ and\
  \citenamefont {Siddiqi}}]{RabiOscillationFeedback}%
  \BibitemOpen
  \bibfield  {author} {\bibinfo {author} {\bibfnamefont {R.}~\bibnamefont
  {Vijay}}, \bibinfo {author} {\bibfnamefont {C.}~\bibnamefont {Macklin}},
  \bibinfo {author} {\bibfnamefont {D.~H.}\ \bibnamefont {Slichter}}, \bibinfo
  {author} {\bibfnamefont {S.~J.}\ \bibnamefont {Weber}}, \bibinfo {author}
  {\bibfnamefont {K.~W.}\ \bibnamefont {Murch}}, \bibinfo {author}
  {\bibfnamefont {R.}~\bibnamefont {Naik}}, \bibinfo {author} {\bibfnamefont
  {A.~N.}\ \bibnamefont {Korotkov}}, \ and\ \bibinfo {author} {\bibfnamefont
  {I.}~\bibnamefont {Siddiqi}},\ }\href {http://dx.doi.org/10.1038/nature11505}
  {\bibfield  {journal} {\bibinfo  {journal} {Nature}\ }\textbf {\bibinfo
  {volume} {490}},\ \bibinfo {pages} {77} (\bibinfo {year} {2012})}\BibitemShut
  {NoStop}%
\bibitem [{\citenamefont {Kubanek}\ \emph {et~al.}(2009)\citenamefont
  {Kubanek}, \citenamefont {Koch}, \citenamefont {Sames}, \citenamefont
  {Ourjoumtsev}, \citenamefont {Pinkse}, \citenamefont {Murr},\ and\
  \citenamefont {Rempe}}]{Photon-by-PhotonFeedback}%
  \BibitemOpen
  \bibfield  {author} {\bibinfo {author} {\bibfnamefont {A.}~\bibnamefont
  {Kubanek}}, \bibinfo {author} {\bibfnamefont {M.}~\bibnamefont {Koch}},
  \bibinfo {author} {\bibfnamefont {C.}~\bibnamefont {Sames}}, \bibinfo
  {author} {\bibfnamefont {A.}~\bibnamefont {Ourjoumtsev}}, \bibinfo {author}
  {\bibfnamefont {P.~W.~H.}\ \bibnamefont {Pinkse}}, \bibinfo {author}
  {\bibfnamefont {K.}~\bibnamefont {Murr}}, \ and\ \bibinfo {author}
  {\bibfnamefont {G.}~\bibnamefont {Rempe}},\ }\href
  {http://dx.doi.org/10.1038/nature08563} {\bibfield  {journal} {\bibinfo
  {journal} {Nature}\ }\textbf {\bibinfo {volume} {462}},\ \bibinfo {pages}
  {898} (\bibinfo {year} {2009})}\BibitemShut {NoStop}%
\bibitem [{\citenamefont {Wilson}\ \emph {et~al.}(2015)\citenamefont {Wilson},
  \citenamefont {Sudhir}, \citenamefont {Piro}, \citenamefont {Schilling},
  \citenamefont {Ghadimi},\ and\ \citenamefont
  {Kippenberg}}]{MechanicalOscillatorFeedback}%
  \BibitemOpen
  \bibfield  {author} {\bibinfo {author} {\bibfnamefont {D.~J.}\ \bibnamefont
  {Wilson}}, \bibinfo {author} {\bibfnamefont {V.}~\bibnamefont {Sudhir}},
  \bibinfo {author} {\bibfnamefont {N.}~\bibnamefont {Piro}}, \bibinfo {author}
  {\bibfnamefont {R.}~\bibnamefont {Schilling}}, \bibinfo {author}
  {\bibfnamefont {A.}~\bibnamefont {Ghadimi}}, \ and\ \bibinfo {author}
  {\bibfnamefont {T.~J.}\ \bibnamefont {Kippenberg}},\ }\href
  {http://dx.doi.org/10.1038/nature14672} {\bibfield  {journal} {\bibinfo
  {journal} {Nature}\ }\textbf {\bibinfo {volume} {524}},\ \bibinfo {pages}
  {325} (\bibinfo {year} {2015})}\BibitemShut {NoStop}%
\bibitem [{\citenamefont {Khalili}\ \emph {et~al.}(2010)\citenamefont
  {Khalili}, \citenamefont {Danilishin}, \citenamefont {Miao}, \citenamefont
  {M\"uller-Ebhardt}, \citenamefont {Yang},\ and\ \citenamefont
  {Chen}}]{YanbeiPreparingaMechanicalOscillator}%
  \BibitemOpen
  \bibfield  {author} {\bibinfo {author} {\bibfnamefont {F.}~\bibnamefont
  {Khalili}}, \bibinfo {author} {\bibfnamefont {S.}~\bibnamefont {Danilishin}},
  \bibinfo {author} {\bibfnamefont {H.}~\bibnamefont {Miao}}, \bibinfo {author}
  {\bibfnamefont {H.}~\bibnamefont {M\"uller-Ebhardt}}, \bibinfo {author}
  {\bibfnamefont {H.}~\bibnamefont {Yang}}, \ and\ \bibinfo {author}
  {\bibfnamefont {Y.}~\bibnamefont {Chen}},\ }\href {\doibase
  10.1103/PhysRevLett.105.070403} {\bibfield  {journal} {\bibinfo  {journal}
  {Phys. Rev. Lett.}\ }\textbf {\bibinfo {volume} {105}},\ \bibinfo {pages}
  {070403} (\bibinfo {year} {2010})}\BibitemShut {NoStop}%
\bibitem [{\citenamefont {Wallquist}\ \emph {et~al.}(2009)\citenamefont
  {Wallquist}, \citenamefont {Hammerer}, \citenamefont {Rabl}, \citenamefont
  {Lukin},\ and\ \citenamefont {Zoller}}]{HybridQuantumDevices}%
  \BibitemOpen
  \bibfield  {author} {\bibinfo {author} {\bibfnamefont {M.}~\bibnamefont
  {Wallquist}}, \bibinfo {author} {\bibfnamefont {K.}~\bibnamefont {Hammerer}},
  \bibinfo {author} {\bibfnamefont {P.}~\bibnamefont {Rabl}}, \bibinfo {author}
  {\bibfnamefont {M.}~\bibnamefont {Lukin}}, \ and\ \bibinfo {author}
  {\bibfnamefont {P.}~\bibnamefont {Zoller}},\ }\href
  {http://stacks.iop.org/1402-4896/2009/i=T137/a=014001} {\bibfield  {journal}
  {\bibinfo  {journal} {Physica Scripta}\ }\textbf {\bibinfo {volume} {T137}},\
  \bibinfo {pages} {014001} (\bibinfo {year} {2009})}\BibitemShut {NoStop}%
\bibitem [{\citenamefont {Gottesman}\ \emph {et~al.}(2001)\citenamefont
  {Gottesman}, \citenamefont {Kitaev},\ and\ \citenamefont {Preskill}}]{GKP}%
  \BibitemOpen
  \bibfield  {author} {\bibinfo {author} {\bibfnamefont {D.}~\bibnamefont
  {Gottesman}}, \bibinfo {author} {\bibfnamefont {A.}~\bibnamefont {Kitaev}}, \
  and\ \bibinfo {author} {\bibfnamefont {J.}~\bibnamefont {Preskill}},\ }\href
  {\doibase 10.1103/PhysRevA.64.012310} {\bibfield  {journal} {\bibinfo
  {journal} {Phys. Rev. A}\ }\textbf {\bibinfo {volume} {64}},\ \bibinfo
  {pages} {012310} (\bibinfo {year} {2001})}\BibitemShut {NoStop}%
\bibitem [{\citenamefont {Menicucci}(2014)}]{MenicucciFaultTolarant}%
  \BibitemOpen
  \bibfield  {author} {\bibinfo {author} {\bibfnamefont {N.~C.}\ \bibnamefont
  {Menicucci}},\ }\href {\doibase 10.1103/PhysRevLett.112.120504} {\bibfield
  {journal} {\bibinfo  {journal} {Phys. Rev. Lett.}\ }\textbf {\bibinfo
  {volume} {112}},\ \bibinfo {pages} {120504} (\bibinfo {year}
  {2014})}\BibitemShut {NoStop}%
\bibitem [{\citenamefont {Furusawa}\ and\ \citenamefont {van
  Loock}(2011)}]{FurusawaBook}%
  \BibitemOpen
  \bibfield  {author} {\bibinfo {author} {\bibfnamefont {A.}~\bibnamefont
  {Furusawa}}\ and\ \bibinfo {author} {\bibfnamefont {P.}~\bibnamefont {van
  Loock}},\ }\href@noop {} {\emph {\bibinfo {title} {Quantum Teleportation and
  Entanglement: A Hybrid Approach to Optical Quantum Information Processing}}}\
  (\bibinfo  {publisher} {Wiley-VCH},\ \bibinfo {year} {2011})\BibitemShut
  {NoStop}%
\bibitem [{\citenamefont {Andersen}\ \emph {et~al.}(2015)\citenamefont
  {Andersen}, \citenamefont {Neergaard-Nielsen}, \citenamefont {van Loock},\
  and\ \citenamefont {Furusawa}}]{UlrikHybrid}%
  \BibitemOpen
  \bibfield  {author} {\bibinfo {author} {\bibfnamefont {U.~L.}\ \bibnamefont
  {Andersen}}, \bibinfo {author} {\bibfnamefont {J.~S.}\ \bibnamefont
  {Neergaard-Nielsen}}, \bibinfo {author} {\bibfnamefont {P.}~\bibnamefont {van
  Loock}}, \ and\ \bibinfo {author} {\bibfnamefont {A.}~\bibnamefont
  {Furusawa}},\ }\href {http://dx.doi.org/10.1038/nphys3410} {\bibfield
  {journal} {\bibinfo  {journal} {Nat Phys}\ }\textbf {\bibinfo {volume}
  {11}},\ \bibinfo {pages} {713} (\bibinfo {year} {2015})}\BibitemShut
  {NoStop}%
\bibitem [{\citenamefont {Kwon}\ and\ \citenamefont
  {Jeong}(2013)}]{BellTestHybrid}%
  \BibitemOpen
  \bibfield  {author} {\bibinfo {author} {\bibfnamefont {H.}~\bibnamefont
  {Kwon}}\ and\ \bibinfo {author} {\bibfnamefont {H.}~\bibnamefont {Jeong}},\
  }\href {\doibase 10.1103/PhysRevA.88.052127} {\bibfield  {journal} {\bibinfo
  {journal} {Phys. Rev. A}\ }\textbf {\bibinfo {volume} {88}},\ \bibinfo
  {pages} {052127} (\bibinfo {year} {2013})}\BibitemShut {NoStop}%
\bibitem [{\citenamefont {Jeong}\ \emph {et~al.}(2014)\citenamefont {Jeong},
  \citenamefont {Zavatta}, \citenamefont {Kang}, \citenamefont {Lee},
  \citenamefont {Costanzo}, \citenamefont {Grandi}, \citenamefont {Ralph},\
  and\ \citenamefont {Bellini}}]{BelliniHybridEntanglement}%
  \BibitemOpen
  \bibfield  {author} {\bibinfo {author} {\bibfnamefont {H.}~\bibnamefont
  {Jeong}}, \bibinfo {author} {\bibfnamefont {A.}~\bibnamefont {Zavatta}},
  \bibinfo {author} {\bibfnamefont {M.}~\bibnamefont {Kang}}, \bibinfo {author}
  {\bibfnamefont {S.-W.}\ \bibnamefont {Lee}}, \bibinfo {author} {\bibfnamefont
  {L.~S.}\ \bibnamefont {Costanzo}}, \bibinfo {author} {\bibfnamefont
  {S.}~\bibnamefont {Grandi}}, \bibinfo {author} {\bibfnamefont {T.~C.}\
  \bibnamefont {Ralph}}, \ and\ \bibinfo {author} {\bibfnamefont
  {M.}~\bibnamefont {Bellini}},\ }\href
  {http://dx.doi.org/10.1038/nphoton.2014.136} {\bibfield  {journal} {\bibinfo
  {journal} {Nat Photon}\ }\textbf {\bibinfo {volume} {8}},\ \bibinfo {pages}
  {564} (\bibinfo {year} {2014})}\BibitemShut {NoStop}%
\bibitem [{\citenamefont {Morin}\ \emph {et~al.}(2014)\citenamefont {Morin},
  \citenamefont {Huang}, \citenamefont {Liu}, \citenamefont {Le~Jeannic},
  \citenamefont {Fabre},\ and\ \citenamefont
  {Laurat}}]{JulienHybridEntanglement}%
  \BibitemOpen
  \bibfield  {author} {\bibinfo {author} {\bibfnamefont {O.}~\bibnamefont
  {Morin}}, \bibinfo {author} {\bibfnamefont {K.}~\bibnamefont {Huang}},
  \bibinfo {author} {\bibfnamefont {J.}~\bibnamefont {Liu}}, \bibinfo {author}
  {\bibfnamefont {H.}~\bibnamefont {Le~Jeannic}}, \bibinfo {author}
  {\bibfnamefont {C.}~\bibnamefont {Fabre}}, \ and\ \bibinfo {author}
  {\bibfnamefont {J.}~\bibnamefont {Laurat}},\ }\href
  {http://dx.doi.org/10.1038/nphoton.2014.137} {\bibfield  {journal} {\bibinfo
  {journal} {Nat Photon}\ }\textbf {\bibinfo {volume} {8}},\ \bibinfo {pages}
  {570} (\bibinfo {year} {2014})}\BibitemShut {NoStop}%
\bibitem [{\citenamefont {Takeda}\ \emph {et~al.}(2013)\citenamefont {Takeda},
  \citenamefont {Mizuta}, \citenamefont {Fuwa}, \citenamefont {van Loock},\
  and\ \citenamefont {Furusawa}}]{TakedaQubitTele}%
  \BibitemOpen
  \bibfield  {author} {\bibinfo {author} {\bibfnamefont {S.}~\bibnamefont
  {Takeda}}, \bibinfo {author} {\bibfnamefont {T.}~\bibnamefont {Mizuta}},
  \bibinfo {author} {\bibfnamefont {M.}~\bibnamefont {Fuwa}}, \bibinfo {author}
  {\bibfnamefont {P.}~\bibnamefont {van Loock}}, \ and\ \bibinfo {author}
  {\bibfnamefont {A.}~\bibnamefont {Furusawa}},\ }\href
  {http://dx.doi.org/10.1038/nature12366} {\bibfield  {journal} {\bibinfo
  {journal} {Nature}\ }\textbf {\bibinfo {volume} {500}},\ \bibinfo {pages}
  {315} (\bibinfo {year} {2013})}\BibitemShut {NoStop}%
\bibitem [{\citenamefont {Miwa}\ \emph {et~al.}(2014)\citenamefont {Miwa},
  \citenamefont {Yoshikawa}, \citenamefont {Iwata}, \citenamefont {Endo},
  \citenamefont {Marek}, \citenamefont {Filip}, \citenamefont {van Loock},\
  and\ \citenamefont {Furusawa}}]{SqueezingSinglePhoton}%
  \BibitemOpen
  \bibfield  {author} {\bibinfo {author} {\bibfnamefont {Y.}~\bibnamefont
  {Miwa}}, \bibinfo {author} {\bibfnamefont {J.}~\bibnamefont {Yoshikawa}},
  \bibinfo {author} {\bibfnamefont {N.}~\bibnamefont {Iwata}}, \bibinfo
  {author} {\bibfnamefont {M.}~\bibnamefont {Endo}}, \bibinfo {author}
  {\bibfnamefont {P.}~\bibnamefont {Marek}}, \bibinfo {author} {\bibfnamefont
  {R.}~\bibnamefont {Filip}}, \bibinfo {author} {\bibfnamefont
  {P.}~\bibnamefont {van Loock}}, \ and\ \bibinfo {author} {\bibfnamefont
  {A.}~\bibnamefont {Furusawa}},\ }\href {\doibase
  10.1103/PhysRevLett.113.013601} {\bibfield  {journal} {\bibinfo  {journal}
  {Phys. Rev. Lett.}\ }\textbf {\bibinfo {volume} {113}},\ \bibinfo {pages}
  {013601} (\bibinfo {year} {2014})}\BibitemShut {NoStop}%
\bibitem [{\citenamefont {Lloyd}\ and\ \citenamefont
  {Braunstein}(1999)}]{QCoverCV}%
  \BibitemOpen
  \bibfield  {author} {\bibinfo {author} {\bibfnamefont {S.}~\bibnamefont
  {Lloyd}}\ and\ \bibinfo {author} {\bibfnamefont {S.~L.}\ \bibnamefont
  {Braunstein}},\ }\href {\doibase 10.1103/PhysRevLett.82.1784} {\bibfield
  {journal} {\bibinfo  {journal} {Phys. Rev. Lett.}\ }\textbf {\bibinfo
  {volume} {82}},\ \bibinfo {pages} {1784} (\bibinfo {year}
  {1999})}\BibitemShut {NoStop}%
\bibitem [{\citenamefont {Bartlett}\ and\ \citenamefont
  {Sanders}(2002)}]{UniversalCVQC}%
  \BibitemOpen
  \bibfield  {author} {\bibinfo {author} {\bibfnamefont {S.~D.}\ \bibnamefont
  {Bartlett}}\ and\ \bibinfo {author} {\bibfnamefont {B.~C.}\ \bibnamefont
  {Sanders}},\ }\href {\doibase 10.1103/PhysRevA.65.042304} {\bibfield
  {journal} {\bibinfo  {journal} {Phys. Rev. A}\ }\textbf {\bibinfo {volume}
  {65}},\ \bibinfo {pages} {042304} (\bibinfo {year} {2002})}\BibitemShut
  {NoStop}%
\bibitem [{\citenamefont {Marek}\ \emph {et~al.}(2011)\citenamefont {Marek},
  \citenamefont {Filip},\ and\ \citenamefont
  {Furusawa}}]{PetrCubicNonlinearity}%
  \BibitemOpen
  \bibfield  {author} {\bibinfo {author} {\bibfnamefont {P.}~\bibnamefont
  {Marek}}, \bibinfo {author} {\bibfnamefont {R.}~\bibnamefont {Filip}}, \ and\
  \bibinfo {author} {\bibfnamefont {A.}~\bibnamefont {Furusawa}},\ }\href
  {\doibase 10.1103/PhysRevA.84.053802} {\bibfield  {journal} {\bibinfo
  {journal} {Phys. Rev. A}\ }\textbf {\bibinfo {volume} {84}},\ \bibinfo
  {pages} {053802} (\bibinfo {year} {2011})}\BibitemShut {NoStop}%
\bibitem [{\citenamefont {Miyata}\ \emph {et~al.}(2014)\citenamefont {Miyata},
  \citenamefont {Ogawa}, \citenamefont {Marek}, \citenamefont {Filip},
  \citenamefont {Yonezawa}, \citenamefont {Yoshikawa},\ and\ \citenamefont
  {Furusawa}}]{DSG}%
  \BibitemOpen
  \bibfield  {author} {\bibinfo {author} {\bibfnamefont {K.}~\bibnamefont
  {Miyata}}, \bibinfo {author} {\bibfnamefont {H.}~\bibnamefont {Ogawa}},
  \bibinfo {author} {\bibfnamefont {P.}~\bibnamefont {Marek}}, \bibinfo
  {author} {\bibfnamefont {R.}~\bibnamefont {Filip}}, \bibinfo {author}
  {\bibfnamefont {H.}~\bibnamefont {Yonezawa}}, \bibinfo {author}
  {\bibfnamefont {J.}~\bibnamefont {Yoshikawa}}, \ and\ \bibinfo {author}
  {\bibfnamefont {A.}~\bibnamefont {Furusawa}},\ }\href {\doibase
  10.1103/PhysRevA.90.060302} {\bibfield  {journal} {\bibinfo  {journal} {Phys.
  Rev. A}\ }\textbf {\bibinfo {volume} {90}},\ \bibinfo {pages} {060302(R)}
  (\bibinfo {year} {2014})}\BibitemShut {NoStop}%
\bibitem [{\citenamefont {Stobi{\'n}ska}\ \emph {et~al.}(2009)\citenamefont
  {Stobi{\'n}ska}, \citenamefont {Alber},\ and\ \citenamefont
  {Leuchs}}]{PerfectExcitation}%
  \BibitemOpen
  \bibfield  {author} {\bibinfo {author} {\bibfnamefont {M.}~\bibnamefont
  {Stobi{\'n}ska}}, \bibinfo {author} {\bibfnamefont {G.}~\bibnamefont
  {Alber}}, \ and\ \bibinfo {author} {\bibfnamefont {G.}~\bibnamefont
  {Leuchs}},\ }\href {http://stacks.iop.org/0295-5075/86/i=1/a=14007}
  {\bibfield  {journal} {\bibinfo  {journal} {EPL (Europhysics Letters)}\
  }\textbf {\bibinfo {volume} {86}},\ \bibinfo {pages} {14007} (\bibinfo {year}
  {2009})}\BibitemShut {NoStop}%
\bibitem [{\citenamefont {Yamamoto}\ and\ \citenamefont
  {James}(2014)}]{ZeroDynamics}%
  \BibitemOpen
  \bibfield  {author} {\bibinfo {author} {\bibfnamefont {N.}~\bibnamefont
  {Yamamoto}}\ and\ \bibinfo {author} {\bibfnamefont {M.~R.}\ \bibnamefont
  {James}},\ }\href {http://stacks.iop.org/1367-2630/16/i=7/a=073032}
  {\bibfield  {journal} {\bibinfo  {journal} {New Journal of Physics}\ }\textbf
  {\bibinfo {volume} {16}},\ \bibinfo {pages} {073032} (\bibinfo {year}
  {2014})}\BibitemShut {NoStop}%
\bibitem [{\citenamefont {Wenner}\ \emph {et~al.}(2014)\citenamefont {Wenner},
  \citenamefont {Yin}, \citenamefont {Chen}, \citenamefont {Barends},
  \citenamefont {Chiaro}, \citenamefont {Jeffrey}, \citenamefont {Kelly},
  \citenamefont {Megrant}, \citenamefont {Mutus}, \citenamefont {Neill},
  \citenamefont {O'Malley}, \citenamefont {Roushan}, \citenamefont {Sank},
  \citenamefont {Vainsencher}, \citenamefont {White}, \citenamefont {Korotkov},
  \citenamefont {Cleland},\ and\ \citenamefont
  {Martinis}}]{CatchingTime-ReversedMicrowave}%
  \BibitemOpen
  \bibfield  {author} {\bibinfo {author} {\bibfnamefont {J.}~\bibnamefont
  {Wenner}}, \bibinfo {author} {\bibfnamefont {Y.}~\bibnamefont {Yin}},
  \bibinfo {author} {\bibfnamefont {Y.}~\bibnamefont {Chen}}, \bibinfo {author}
  {\bibfnamefont {R.}~\bibnamefont {Barends}}, \bibinfo {author} {\bibfnamefont
  {B.}~\bibnamefont {Chiaro}}, \bibinfo {author} {\bibfnamefont
  {E.}~\bibnamefont {Jeffrey}}, \bibinfo {author} {\bibfnamefont
  {J.}~\bibnamefont {Kelly}}, \bibinfo {author} {\bibfnamefont
  {A.}~\bibnamefont {Megrant}}, \bibinfo {author} {\bibfnamefont {J.~Y.}\
  \bibnamefont {Mutus}}, \bibinfo {author} {\bibfnamefont {C.}~\bibnamefont
  {Neill}}, \bibinfo {author} {\bibfnamefont {P.~J.~J.}\ \bibnamefont
  {O'Malley}}, \bibinfo {author} {\bibfnamefont {P.}~\bibnamefont {Roushan}},
  \bibinfo {author} {\bibfnamefont {D.}~\bibnamefont {Sank}}, \bibinfo {author}
  {\bibfnamefont {A.}~\bibnamefont {Vainsencher}}, \bibinfo {author}
  {\bibfnamefont {T.~C.}\ \bibnamefont {White}}, \bibinfo {author}
  {\bibfnamefont {A.~N.}\ \bibnamefont {Korotkov}}, \bibinfo {author}
  {\bibfnamefont {A.~N.}\ \bibnamefont {Cleland}}, \ and\ \bibinfo {author}
  {\bibfnamefont {J.~M.}\ \bibnamefont {Martinis}},\ }\href {\doibase
  10.1103/PhysRevLett.112.210501} {\bibfield  {journal} {\bibinfo  {journal}
  {Phys. Rev. Lett.}\ }\textbf {\bibinfo {volume} {112}},\ \bibinfo {pages}
  {210501} (\bibinfo {year} {2014})}\BibitemShut {NoStop}%
\bibitem [{\citenamefont {Cirac}\ \emph {et~al.}(1997)\citenamefont {Cirac},
  \citenamefont {Zoller}, \citenamefont {Kimble},\ and\ \citenamefont
  {Mabuchi}}]{QuantumStateTransfer}%
  \BibitemOpen
  \bibfield  {author} {\bibinfo {author} {\bibfnamefont {J.~I.}\ \bibnamefont
  {Cirac}}, \bibinfo {author} {\bibfnamefont {P.}~\bibnamefont {Zoller}},
  \bibinfo {author} {\bibfnamefont {H.~J.}\ \bibnamefont {Kimble}}, \ and\
  \bibinfo {author} {\bibfnamefont {H.}~\bibnamefont {Mabuchi}},\ }\href
  {\doibase 10.1103/PhysRevLett.78.3221} {\bibfield  {journal} {\bibinfo
  {journal} {Phys. Rev. Lett.}\ }\textbf {\bibinfo {volume} {78}},\ \bibinfo
  {pages} {3221} (\bibinfo {year} {1997})}\BibitemShut {NoStop}%
\bibitem [{\citenamefont {Kimble}(2008)}]{KimbleQuantumInternet}%
  \BibitemOpen
  \bibfield  {author} {\bibinfo {author} {\bibfnamefont {H.~J.}\ \bibnamefont
  {Kimble}},\ }\href {http://dx.doi.org/10.1038/nature07127} {\bibfield
  {journal} {\bibinfo  {journal} {Nature}\ }\textbf {\bibinfo {volume} {453}},\
  \bibinfo {pages} {1023} (\bibinfo {year} {2008})}\BibitemShut {NoStop}%
\bibitem [{\citenamefont {Duan}\ \emph {et~al.}(2001)\citenamefont {Duan},
  \citenamefont {Lukin}, \citenamefont {Cirac},\ and\ \citenamefont
  {Zoller}}]{DLCZ}%
  \BibitemOpen
  \bibfield  {author} {\bibinfo {author} {\bibfnamefont {L.~M.}\ \bibnamefont
  {Duan}}, \bibinfo {author} {\bibfnamefont {M.~D.}\ \bibnamefont {Lukin}},
  \bibinfo {author} {\bibfnamefont {J.~I.}\ \bibnamefont {Cirac}}, \ and\
  \bibinfo {author} {\bibfnamefont {P.}~\bibnamefont {Zoller}},\ }\href
  {http://dx.doi.org/10.1038/35106500} {\bibfield  {journal} {\bibinfo
  {journal} {Nature}\ }\textbf {\bibinfo {volume} {414}},\ \bibinfo {pages}
  {413} (\bibinfo {year} {2001})}\BibitemShut {NoStop}%
\bibitem [{\citenamefont {Lvovsky}\ \emph {et~al.}(2001)\citenamefont
  {Lvovsky}, \citenamefont {Hansen}, \citenamefont {Aichele}, \citenamefont
  {Benson}, \citenamefont {Mlynek},\ and\ \citenamefont
  {Schiller}}]{LvovskySinglePhotonHomodyne}%
  \BibitemOpen
  \bibfield  {author} {\bibinfo {author} {\bibfnamefont {A.~I.}\ \bibnamefont
  {Lvovsky}}, \bibinfo {author} {\bibfnamefont {H.}~\bibnamefont {Hansen}},
  \bibinfo {author} {\bibfnamefont {T.}~\bibnamefont {Aichele}}, \bibinfo
  {author} {\bibfnamefont {O.}~\bibnamefont {Benson}}, \bibinfo {author}
  {\bibfnamefont {J.}~\bibnamefont {Mlynek}}, \ and\ \bibinfo {author}
  {\bibfnamefont {S.}~\bibnamefont {Schiller}},\ }\href {\doibase
  10.1103/PhysRevLett.87.050402} {\bibfield  {journal} {\bibinfo  {journal}
  {Phys. Rev. Lett.}\ }\textbf {\bibinfo {volume} {87}},\ \bibinfo {pages}
  {050402} (\bibinfo {year} {2001})}\BibitemShut {NoStop}%
\bibitem [{\citenamefont {Polycarpou}\ \emph {et~al.}(2012)\citenamefont
  {Polycarpou}, \citenamefont {Cassemiro}, \citenamefont {Venturi},
  \citenamefont {Zavatta},\ and\ \citenamefont
  {Bellini}}]{BelliniAdaptiveDetection}%
  \BibitemOpen
  \bibfield  {author} {\bibinfo {author} {\bibfnamefont {C.}~\bibnamefont
  {Polycarpou}}, \bibinfo {author} {\bibfnamefont {K.~N.}\ \bibnamefont
  {Cassemiro}}, \bibinfo {author} {\bibfnamefont {G.}~\bibnamefont {Venturi}},
  \bibinfo {author} {\bibfnamefont {A.}~\bibnamefont {Zavatta}}, \ and\
  \bibinfo {author} {\bibfnamefont {M.}~\bibnamefont {Bellini}},\ }\href
  {\doibase 10.1103/PhysRevLett.109.053602} {\bibfield  {journal} {\bibinfo
  {journal} {Phys. Rev. Lett.}\ }\textbf {\bibinfo {volume} {109}},\ \bibinfo
  {pages} {053602} (\bibinfo {year} {2012})}\BibitemShut {NoStop}%
\bibitem [{\citenamefont {Neergaard-Nielsen}\ \emph {et~al.}(2006)\citenamefont
  {Neergaard-Nielsen}, \citenamefont {Nielsen}, \citenamefont {Hettich},
  \citenamefont {M\o{}lmer},\ and\ \citenamefont {Polzik}}]{JonasCat}%
  \BibitemOpen
  \bibfield  {author} {\bibinfo {author} {\bibfnamefont {J.~S.}\ \bibnamefont
  {Neergaard-Nielsen}}, \bibinfo {author} {\bibfnamefont {B.~M.}\ \bibnamefont
  {Nielsen}}, \bibinfo {author} {\bibfnamefont {C.}~\bibnamefont {Hettich}},
  \bibinfo {author} {\bibfnamefont {K.}~\bibnamefont {M\o{}lmer}}, \ and\
  \bibinfo {author} {\bibfnamefont {E.~S.}\ \bibnamefont {Polzik}},\ }\href
  {\doibase 10.1103/PhysRevLett.97.083604} {\bibfield  {journal} {\bibinfo
  {journal} {Phys. Rev. Lett.}\ }\textbf {\bibinfo {volume} {97}},\ \bibinfo
  {pages} {083604} (\bibinfo {year} {2006})}\BibitemShut {NoStop}%
\bibitem [{\citenamefont {Neergaard-Nielsen}\ \emph {et~al.}(2007)\citenamefont
  {Neergaard-Nielsen}, \citenamefont {Nielsen}, \citenamefont {Takahashi},
  \citenamefont {Vistnes},\ and\ \citenamefont
  {Polzik}}]{JonasBrightSinglePhoton}%
  \BibitemOpen
  \bibfield  {author} {\bibinfo {author} {\bibfnamefont {J.~S.}\ \bibnamefont
  {Neergaard-Nielsen}}, \bibinfo {author} {\bibfnamefont {B.~M.}\ \bibnamefont
  {Nielsen}}, \bibinfo {author} {\bibfnamefont {H.}~\bibnamefont {Takahashi}},
  \bibinfo {author} {\bibfnamefont {A.~I.}\ \bibnamefont {Vistnes}}, \ and\
  \bibinfo {author} {\bibfnamefont {E.~S.}\ \bibnamefont {Polzik}},\ }\href
  {\doibase 10.1364/OE.15.007940} {\bibfield  {journal} {\bibinfo  {journal}
  {Opt. Express}\ }\textbf {\bibinfo {volume} {15}},\ \bibinfo {pages} {7940}
  (\bibinfo {year} {2007})}\BibitemShut {NoStop}%
\bibitem [{\citenamefont {MacRae}\ \emph {et~al.}(2012)\citenamefont {MacRae},
  \citenamefont {Brannan}, \citenamefont {Achal},\ and\ \citenamefont
  {Lvovsky}}]{LvovskyNarrowBandPhoton}%
  \BibitemOpen
  \bibfield  {author} {\bibinfo {author} {\bibfnamefont {A.}~\bibnamefont
  {MacRae}}, \bibinfo {author} {\bibfnamefont {T.}~\bibnamefont {Brannan}},
  \bibinfo {author} {\bibfnamefont {R.}~\bibnamefont {Achal}}, \ and\ \bibinfo
  {author} {\bibfnamefont {A.~I.}\ \bibnamefont {Lvovsky}},\ }\href {\doibase
  10.1103/PhysRevLett.109.033601} {\bibfield  {journal} {\bibinfo  {journal}
  {Phys. Rev. Lett.}\ }\textbf {\bibinfo {volume} {109}},\ \bibinfo {pages}
  {033601} (\bibinfo {year} {2012})}\BibitemShut {NoStop}%
\bibitem [{\citenamefont {Lee}\ \emph {et~al.}(2011)\citenamefont {Lee},
  \citenamefont {Benichi}, \citenamefont {Takeno}, \citenamefont {Takeda},
  \citenamefont {Webb}, \citenamefont {Huntington},\ and\ \citenamefont
  {Furusawa}}]{LeeCatTele}%
  \BibitemOpen
  \bibfield  {author} {\bibinfo {author} {\bibfnamefont {N.}~\bibnamefont
  {Lee}}, \bibinfo {author} {\bibfnamefont {H.}~\bibnamefont {Benichi}},
  \bibinfo {author} {\bibfnamefont {Y.}~\bibnamefont {Takeno}}, \bibinfo
  {author} {\bibfnamefont {S.}~\bibnamefont {Takeda}}, \bibinfo {author}
  {\bibfnamefont {J.}~\bibnamefont {Webb}}, \bibinfo {author} {\bibfnamefont
  {E.}~\bibnamefont {Huntington}}, \ and\ \bibinfo {author} {\bibfnamefont
  {A.}~\bibnamefont {Furusawa}},\ }\href {\doibase 10.1126/science.1201034}
  {\bibfield  {journal} {\bibinfo  {journal} {Science}\ }\textbf {\bibinfo
  {volume} {332}},\ \bibinfo {pages} {330} (\bibinfo {year}
  {2011})}\BibitemShut {NoStop}%
\bibitem [{\citenamefont {Ou}\ and\ \citenamefont
  {Kimble}(1995)}]{KimblePhotoDetectionProcesses}%
  \BibitemOpen
  \bibfield  {author} {\bibinfo {author} {\bibfnamefont {Z.~Y.}\ \bibnamefont
  {Ou}}\ and\ \bibinfo {author} {\bibfnamefont {H.~J.}\ \bibnamefont
  {Kimble}},\ }\href {\doibase 10.1103/PhysRevA.52.3126} {\bibfield  {journal}
  {\bibinfo  {journal} {Phys. Rev. A}\ }\textbf {\bibinfo {volume} {52}},\
  \bibinfo {pages} {3126} (\bibinfo {year} {1995})}\BibitemShut {NoStop}%
\bibitem [{\citenamefont {Nielsen}\ and\ \citenamefont
  {M\o{}lmer}(2007)}]{NielsenSinglePhoton}%
  \BibitemOpen
  \bibfield  {author} {\bibinfo {author} {\bibfnamefont {A.~E.~B.}\
  \bibnamefont {Nielsen}}\ and\ \bibinfo {author} {\bibfnamefont
  {K.}~\bibnamefont {M\o{}lmer}},\ }\href {\doibase 10.1103/PhysRevA.75.023806}
  {\bibfield  {journal} {\bibinfo  {journal} {Phys. Rev. A}\ }\textbf {\bibinfo
  {volume} {75}},\ \bibinfo {pages} {023806} (\bibinfo {year}
  {2007})}\BibitemShut {NoStop}%
\bibitem [{\citenamefont {Qin}\ \emph {et~al.}(2015)\citenamefont {Qin},
  \citenamefont {Prasad}, \citenamefont {Brannan}, \citenamefont {MacRae},
  \citenamefont {Lezama},\ and\ \citenamefont {Lvovsky}}]{LvovskyComplete}%
  \BibitemOpen
  \bibfield  {author} {\bibinfo {author} {\bibfnamefont {Z.}~\bibnamefont
  {Qin}}, \bibinfo {author} {\bibfnamefont {A.~S.}\ \bibnamefont {Prasad}},
  \bibinfo {author} {\bibfnamefont {T.}~\bibnamefont {Brannan}}, \bibinfo
  {author} {\bibfnamefont {A.}~\bibnamefont {MacRae}}, \bibinfo {author}
  {\bibfnamefont {A.}~\bibnamefont {Lezama}}, \ and\ \bibinfo {author}
  {\bibfnamefont {A.~I.}\ \bibnamefont {Lvovsky}},\ }\href
  {http://dx.doi.org/10.1038/lsa.2015.71} {\bibfield  {journal} {\bibinfo
  {journal} {Light Sci Appl}\ }\textbf {\bibinfo {volume} {4}},\ \bibinfo
  {pages} {e298} (\bibinfo {year} {2015})}\BibitemShut {NoStop}%
\bibitem [{\citenamefont {Zhang}\ \emph {et~al.}(2012)\citenamefont {Zhang},
  \citenamefont {Liu}, \citenamefont {Zhou}, \citenamefont {Chuu},
  \citenamefont {Loy},\ and\ \citenamefont {Du}}]{CoherentControl}%
  \BibitemOpen
  \bibfield  {author} {\bibinfo {author} {\bibfnamefont {S.}~\bibnamefont
  {Zhang}}, \bibinfo {author} {\bibfnamefont {C.}~\bibnamefont {Liu}}, \bibinfo
  {author} {\bibfnamefont {S.}~\bibnamefont {Zhou}}, \bibinfo {author}
  {\bibfnamefont {C.-S.}\ \bibnamefont {Chuu}}, \bibinfo {author}
  {\bibfnamefont {M.~M.~T.}\ \bibnamefont {Loy}}, \ and\ \bibinfo {author}
  {\bibfnamefont {S.}~\bibnamefont {Du}},\ }\href {\doibase
  10.1103/PhysRevLett.109.263601} {\bibfield  {journal} {\bibinfo  {journal}
  {Phys. Rev. Lett.}\ }\textbf {\bibinfo {volume} {109}},\ \bibinfo {pages}
  {263601} (\bibinfo {year} {2012})}\BibitemShut {NoStop}%
\bibitem [{\citenamefont {Gulati}\ \emph {et~al.}(2014)\citenamefont {Gulati},
  \citenamefont {Srivathsan}, \citenamefont {Chng}, \citenamefont {Cer\`e},
  \citenamefont {Matsukevich},\ and\ \citenamefont
  {Kurtsiefer}}]{GulatiGenerationExprise}%
  \BibitemOpen
  \bibfield  {author} {\bibinfo {author} {\bibfnamefont {G.~K.}\ \bibnamefont
  {Gulati}}, \bibinfo {author} {\bibfnamefont {B.}~\bibnamefont {Srivathsan}},
  \bibinfo {author} {\bibfnamefont {B.}~\bibnamefont {Chng}}, \bibinfo {author}
  {\bibfnamefont {A.}~\bibnamefont {Cer\`e}}, \bibinfo {author} {\bibfnamefont
  {D.}~\bibnamefont {Matsukevich}}, \ and\ \bibinfo {author} {\bibfnamefont
  {C.}~\bibnamefont {Kurtsiefer}},\ }\href {\doibase
  10.1103/PhysRevA.90.033819} {\bibfield  {journal} {\bibinfo  {journal} {Phys.
  Rev. A}\ }\textbf {\bibinfo {volume} {90}},\ \bibinfo {pages} {033819}
  (\bibinfo {year} {2014})}\BibitemShut {NoStop}%
\bibitem [{\citenamefont {Liu}\ \emph {et~al.}(2014)\citenamefont {Liu},
  \citenamefont {Sun}, \citenamefont {Zhao}, \citenamefont {Zhang},
  \citenamefont {Loy},\ and\ \citenamefont {Du}}]{EfficientlyLoading}%
  \BibitemOpen
  \bibfield  {author} {\bibinfo {author} {\bibfnamefont {C.}~\bibnamefont
  {Liu}}, \bibinfo {author} {\bibfnamefont {Y.}~\bibnamefont {Sun}}, \bibinfo
  {author} {\bibfnamefont {L.}~\bibnamefont {Zhao}}, \bibinfo {author}
  {\bibfnamefont {S.}~\bibnamefont {Zhang}}, \bibinfo {author} {\bibfnamefont
  {M.~M.~T.}\ \bibnamefont {Loy}}, \ and\ \bibinfo {author} {\bibfnamefont
  {S.}~\bibnamefont {Du}},\ }\href {\doibase 10.1103/PhysRevLett.113.133601}
  {\bibfield  {journal} {\bibinfo  {journal} {Phys. Rev. Lett.}\ }\textbf
  {\bibinfo {volume} {113}},\ \bibinfo {pages} {133601} (\bibinfo {year}
  {2014})}\BibitemShut {NoStop}%
\bibitem [{\citenamefont {Srivathsan}\ \emph {et~al.}(2014)\citenamefont
  {Srivathsan}, \citenamefont {Gulati}, \citenamefont {Cer\`e}, \citenamefont
  {Chng},\ and\ \citenamefont {Kurtsiefer}}]{Reversing}%
  \BibitemOpen
  \bibfield  {author} {\bibinfo {author} {\bibfnamefont {B.}~\bibnamefont
  {Srivathsan}}, \bibinfo {author} {\bibfnamefont {G.~K.}\ \bibnamefont
  {Gulati}}, \bibinfo {author} {\bibfnamefont {A.}~\bibnamefont {Cer\`e}},
  \bibinfo {author} {\bibfnamefont {B.}~\bibnamefont {Chng}}, \ and\ \bibinfo
  {author} {\bibfnamefont {C.}~\bibnamefont {Kurtsiefer}},\ }\href {\doibase
  10.1103/PhysRevLett.113.163601} {\bibfield  {journal} {\bibinfo  {journal}
  {Phys. Rev. Lett.}\ }\textbf {\bibinfo {volume} {113}},\ \bibinfo {pages}
  {163601} (\bibinfo {year} {2014})}\BibitemShut {NoStop}%
\bibitem [{\citenamefont {Bimbard}\ \emph {et~al.}(2010)\citenamefont
  {Bimbard}, \citenamefont {Jain}, \citenamefont {MacRae},\ and\ \citenamefont
  {Lvovsky}}]{Lvovsky2photon}%
  \BibitemOpen
  \bibfield  {author} {\bibinfo {author} {\bibfnamefont {E.}~\bibnamefont
  {Bimbard}}, \bibinfo {author} {\bibfnamefont {N.}~\bibnamefont {Jain}},
  \bibinfo {author} {\bibfnamefont {A.}~\bibnamefont {MacRae}}, \ and\ \bibinfo
  {author} {\bibfnamefont {A.~I.}\ \bibnamefont {Lvovsky}},\ }\href
  {http://dx.doi.org/10.1038/nphoton.2010.6} {\bibfield  {journal} {\bibinfo
  {journal} {Nat Photon}\ }\textbf {\bibinfo {volume} {4}},\ \bibinfo {pages}
  {243} (\bibinfo {year} {2010})}\BibitemShut {NoStop}%
\bibitem [{\citenamefont {Yukawa}\ \emph {et~al.}(2013)\citenamefont {Yukawa},
  \citenamefont {Miyata}, \citenamefont {Mizuta}, \citenamefont {Yonezawa},
  \citenamefont {Marek}, \citenamefont {Filip},\ and\ \citenamefont
  {Furusawa}}]{YukawaSuperposition}%
  \BibitemOpen
  \bibfield  {author} {\bibinfo {author} {\bibfnamefont {M.}~\bibnamefont
  {Yukawa}}, \bibinfo {author} {\bibfnamefont {K.}~\bibnamefont {Miyata}},
  \bibinfo {author} {\bibfnamefont {T.}~\bibnamefont {Mizuta}}, \bibinfo
  {author} {\bibfnamefont {H.}~\bibnamefont {Yonezawa}}, \bibinfo {author}
  {\bibfnamefont {P.}~\bibnamefont {Marek}}, \bibinfo {author} {\bibfnamefont
  {R.}~\bibnamefont {Filip}}, \ and\ \bibinfo {author} {\bibfnamefont
  {A.}~\bibnamefont {Furusawa}},\ }\href {\doibase 10.1364/OE.21.005529}
  {\bibfield  {journal} {\bibinfo  {journal} {Opt. Express}\ }\textbf {\bibinfo
  {volume} {21}},\ \bibinfo {pages} {5529} (\bibinfo {year}
  {2013})}\BibitemShut {NoStop}%
\bibitem [{\citenamefont {Morin}\ \emph {et~al.}(2013)\citenamefont {Morin},
  \citenamefont {Fabre},\ and\ \citenamefont {Laurat}}]{JulienTemporalMode}%
  \BibitemOpen
  \bibfield  {author} {\bibinfo {author} {\bibfnamefont {O.}~\bibnamefont
  {Morin}}, \bibinfo {author} {\bibfnamefont {C.}~\bibnamefont {Fabre}}, \ and\
  \bibinfo {author} {\bibfnamefont {J.}~\bibnamefont {Laurat}},\ }\href
  {\doibase 10.1103/PhysRevLett.111.213602} {\bibfield  {journal} {\bibinfo
  {journal} {Phys. Rev. Lett.}\ }\textbf {\bibinfo {volume} {111}},\ \bibinfo
  {pages} {213602} (\bibinfo {year} {2013})}\BibitemShut {NoStop}%
\bibitem [{sup()}]{supplemental}%
  \BibitemOpen
  \href@noop {} {}\bibinfo {howpublished} {see supplemental material at [] for
  the video}\BibitemShut {NoStop}%
\bibitem [{\citenamefont {Strekalov}\ \emph {et~al.}(1995)\citenamefont
  {Strekalov}, \citenamefont {Sergienko}, \citenamefont {Klyshko},\ and\
  \citenamefont {Shih}}]{TwoPhotonGhost}%
  \BibitemOpen
  \bibfield  {author} {\bibinfo {author} {\bibfnamefont {D.~V.}\ \bibnamefont
  {Strekalov}}, \bibinfo {author} {\bibfnamefont {A.~V.}\ \bibnamefont
  {Sergienko}}, \bibinfo {author} {\bibfnamefont {D.~N.}\ \bibnamefont
  {Klyshko}}, \ and\ \bibinfo {author} {\bibfnamefont {Y.~H.}\ \bibnamefont
  {Shih}},\ }\href {\doibase 10.1103/PhysRevLett.74.3600} {\bibfield  {journal}
  {\bibinfo  {journal} {Phys. Rev. Lett.}\ }\textbf {\bibinfo {volume} {74}},\
  \bibinfo {pages} {3600} (\bibinfo {year} {1995})}\BibitemShut {NoStop}%
\bibitem [{\citenamefont {Banaszek}(1998)}]{Maxlik}%
  \BibitemOpen
  \bibfield  {author} {\bibinfo {author} {\bibfnamefont {K.}~\bibnamefont
  {Banaszek}},\ }\href {\doibase 10.1103/PhysRevA.57.5013} {\bibfield
  {journal} {\bibinfo  {journal} {Phys. Rev. A}\ }\textbf {\bibinfo {volume}
  {57}},\ \bibinfo {pages} {5013} (\bibinfo {year} {1998})}\BibitemShut
  {NoStop}%
\end{thebibliography}%

\end{document}